\documentclass{article}

\usepackage{amssymb,amsfonts,amsmath,stmaryrd}
\usepackage{cite,enumerate,float,indentfirst}
\usepackage{color}
\usepackage{tikz}
\usetikzlibrary{arrows,snakes,backgrounds}

\def\be{\begin{eqnarray}}
\def\ee{\end{eqnarray}}
\def\nn{\nonumber}

\def\p{\partial}
\def\tr{{\rm tr}\,}

\definecolor{red}{rgb}{1,0,0}
\definecolor{orange}{rgb}{1,0.5,0}
\definecolor{violet}{rgb}{0.7,0,1}

\hoffset= - 1.0in         % switch off for draft style
\begin{document}
\title{\bf
Towards elliptic deformation of $q,t$-matrix models
}

\author{
{\bf Andrei Mironov$^{a,b,c}$}\footnote{mironov@lpi.ru; mironov@itep.ru},
\ and  \  {\bf Alexei Morozov$^{d,b,c}$}\thanks{morozov@itep.ru}
}
\date{ }

\maketitle

\vspace{-5.0cm}

\begin{center}
\hfill FIAN/TD-14/20\\
\hfill IITP/TH-17/20\\
\hfill ITEP/TH-26/20\\
\hfill MIPT/TH-15/20
\end{center}

\vspace{2.cm}

\begin{center}
$^a$ {\small {\it Lebedev Physics Institute, Moscow 119991, Russia}}\\
$^b$ {\small {\it ITEP, Moscow 117218, Russia}}\\
$^c$ {\small {\it Institute for Information Transmission Problems, Moscow 127994, Russia}}\\
$^d$ {\small {\it MIPT, Dolgoprudny, 141701, Russia}}
\end{center}

\vspace{.0cm}

\begin{abstract}
As a necessary step in construction of elliptic matrix models,
which preserve the superintegrability property $<char>\sim {\rm char}$,
we suggest an elliptic deformation of the peculiar loci
$p_k^{\Delta_n}$, which play an important role
in precise formulation of this property.
The suggestion is to define the $p_k^{\Delta_n}$-loci as elliptic functions
with the right asymptotics at $\tau\to i\infty$.
If this hypothesis is correct, one can move to substituting the
Schur and Macdonald functions in the role of characters by the
elliptic GNS and study their behavior at the deformed topological
and $\Delta$ loci.
\end{abstract}

\section{Introduction}

Eigenvalue matrix models can be defined in a number of different ways:
through matrix integrals,
through various equations (Virasoro and W-like constraints)
imposed on partition functions,
through recursion relations on averages and so on.
The main observation coming from all these approaches,
is apparent integrability (see \cite{UFN3} for comprehensive reviews).
However, there is actually more than that:
these theories appear to be {\it super}integrable,
i.e. correlators can be evaluated explicitly in terms of
reasonably-elementary functions \cite{HZ}.
Of course, this is true for a cleverly adjusted basis of observables
\cite{MM}, which are usually not the naive multi-trace operators,
but their peculiar combinations described by ``characters" $\chi_R$,
the polynomials from the Schur-Macdonald family,
see \cite{MMsumrules} for various examples.
The averages of characters are again expressed through the polynomials
of the same kind, evaluated at some peculiar loci in the space of
time variables: the topological locus $p_k = N$ where the characters 
reduce to dimensions of various representations, and
$\Delta_{n}$ loci, associated with $p_k = \delta_{k,n}$.
The simplest, still representative example is provided by
the Gaussian Hermitian model where
\be
\left< \chi_R(\tr M^k)\right>_N =
\frac{\chi_R(\delta_{k,2})\chi_R(N)}{\chi_R(\delta_{k,1})}
\label{GHM}
\ee
see \cite{MM} for notation and other details.
Well known are $q$- and $q,t$-deformations of this model \cite{MPSh,MMell},
where Schur polynomials $\chi_R$ are substituted by Macdonald polynomials
${\rm Mac}_R$,
and loci change for
\be
p_k^* = \frac{1-t^{Nk}}{1-t^k} \ \stackrel{t\rightarrow 1}{\longrightarrow} \ N
\ee
and
\be
p_k^{\Delta_n} = \frac{\sum_{j=0}^{n-1}e^{2\pi i{kj\over n}}}{1-t^k}
\ \stackrel{t\rightarrow 1}{\longrightarrow} \ \delta_{k,n}
\label{Delqt}
\ee

In this paper, we make a step towards a further elliptic deformation,
with an additional parameter $e^{i\pi \tau}\neq 0$,
where the Macdonald polynomials are supposed to be substituted by
the elliptic GNS polynomials \cite{AKMMgns1,AKMMgns2}
(generalization of  the Noumi-Shiraishi polynomials from \cite{NS}).
Here we will concentrate on a particular question of what is the
elliptic generalization of (\ref{Delqt}) and what happens on this
generalized locus to Schur polynomials. We demonstrate that a proper choice of the elliptic deformation of loci completely preserves the factorization property of the Schur polynomials.
Lifting to GNS is left for further consideration. An accompanying discussion of the elliptic matrix model itself can be found in \cite{MMell}. In particular, it is explained in \cite{MMell} that the factorization at the elliptic loci already at the level of the Schur functions is required in the elliptic matrix model.

\section{The idea}

A very natural assumption is that $\frac{1}{1-t^{k}}$ in the definition of
the loci $\Delta_n$  are substitued
by elliptic functions, i.e. by rational combinations of
Weierstrass function $X = {\wp}(z)$ and its derivative $Y ={\wp}'(z)$.
One should only care to reproduce the right asymptotics at $z\longrightarrow 1$
and express it through $t = e^{2\pi i z}$.

The main supporting observation is that, on the loci $p_k^{\Delta_n}$,
the Schur functions are factorized:
\be\label{4}
\chi_R\left(p_k=\frac{1}{1-t^{k}}\right) = {t^{\nu_R/2}\over \prod_{i,j\in R}\left(
1-t^{h_{i,j}}\right)}\nn \\
\chi_R\left(p_{k} = \frac{1+(-1)^k}{1-t^{k}}\right) = {{\cal C}_Rt^{\mu_R/2}\over \prod_{i,j\in R:\ h_{i,j}\in
\mathbb{Z}_{2n}}\left(
1-t^{h_{i,j}}\right)}
\ee
Here ${\cal C}_R$ is $+1$ if $\chi_R(\delta_{k,2})>0$, $0$ if $\chi_R(\delta_{k,2})=0$ and $-1$ if $\chi_R(\delta_{k,2})<0$. The exponent $\nu_R:=2\sum_iR_i(i-1)$, and $\mu_R$ is much more complicated exponent (see sec.5). $h_{i,j}$ is the corresponding hook length.
We will see that this remains true after suggested
elliptic deformation.

\section{Trigonometric case}

As a warm-up we can consider trigonometric approximation, where
\be
x:= -\p^2_z \log(\sin\pi z) - \frac{\pi^2}{3} = \frac{\pi ^2}{\sin^2(\pi z)} - \frac{\pi^2}{3}
\nn \\
y: = -\p^3_z \log(\sin\pi z) = \frac{2\pi^3\cos(\pi z)}{\sin^3(\pi z)}
\label{xyvsz}
\ee
satisfy
\be
y^2 = 4x^3 - \frac{4\pi^4}{3}x - \frac{8\pi^6}{27}={4\over 27}\Big(3x-2\pi^2\Big)\Big(3x+\pi^2\Big)^2
\ee
Note that in the second derivative we can substitute $\sin(\pi z)$ by $1-e^{2\pi i z} = 1-t$,
only derivatives should be changed from $\p_z$ to $2\pi i t\p_t$:
\be
\Big((t\p_t)^3\log(1-t)\Big)^2 + 4\left((t\p_t)^2\log(1-t)-\frac{1}{12}\right)^3
- \frac{1}{12}\left((t\p_t)^2\log(1-t)-\frac{1}{12}\right) - \frac{1}{216} = 0
\ee

We need an elliptic deformation which is formulated in terms of the doubly periodic Weierstrass function $\wp(z)$:
\be\label{wp}
X:={\wp}(z) =  -\p^2_z \log\left[\theta\Big({z\over 2\omega}\Big|\tau\Big)\right] - {\eta\over\omega}
\nn \\
\nn\\
Y: = {\wp}'(z)= -\p^3_z \log\left[\theta\Big({z\over 2\omega}\Big|\tau\Big)\right]
\ee
where $\omega$ is one of the periods, $\theta(v|\tau)$ is the Jacobi $\theta$-function, and $\eta$ is a constant, see \cite{BE}. 
Then,
\be
Y^2 = 4X^3 - g_2(\tau) X - g_3(\tau)  
\ee
Since the arguments of $\theta$ in (\ref{wp}) is ${z\over 2\omega}$, it is natural to associate $1-e^{i\pi z/\omega}$ to $1-t$ and $\p_z$ to ${\pi i \over\omega}t\p_t$. Define using the triple product formula for the $\theta$-function (so that the trigonometric limit above is given at $\tau\to i\infty$)
\be
\theta\Big({z\over 2\omega}\Big|\tau\Big)=ie^{i\pi\tau/4}{\sqrt{2\omega}\over\sqrt{z}}(1-e^{i\pi z/\omega})
\prod_{n=1}^\infty\Big(1-e^{2\pi in\tau}e^{i\pi z/\omega}\Big)\Big(1-e^{2\pi in\tau}e^{-i\pi z/\omega}\Big)\Big(1-e^{2\pi in\tau}\Big)=:\vartheta_\tau(t)
\ee
so that finally one gets
\be
\Big((t\p_t)^3\log(\vartheta_\tau(t))\Big)^2 + 4\Big((t\p_t)^2\log(\vartheta_\tau(t))-{\omega\eta\over\pi^2}\Big)^3
- g_2(\tau){\omega^8\over\pi^8} \cdot \Big((t\p_t)^2\log(\vartheta_\tau(t))-{\omega\eta\over\pi^2}\Big) - {\omega^{12}\over\pi^{12}}g_3(\tau) = 0
\ee

\section{Elliptic $\Delta$ loci}

The functions $p_k^\Delta = \frac{1}{1-t^k}$ in the definition of $\Delta_n$ have
first order poles at $t=1$ and thus have no chances to be elliptic functions.
However, the time-variables $p_k$ are actually defined up to a grade-preserving
rescaling $p_k\longrightarrow c^kp_k$, and this can be used to cure the problem.
For example,
\be
p_k^\Delta = \frac{1}{(1-t^k)(1-t)^k}
\ee
or even
\be\label{xy}
p_k^\Delta = \frac{(1+t)^{2k}}{(1-t^k)(1-t)^k}
\ee
have poles of order $k+1\geq 2$ at $t=1$ and
can possess simple elliptic continuations.

Such a continuation is not unique, but defined up to a triangular redefinition.
However, it makes sense to begin with the most natural try,
when $p_1 \stackrel{?}{=} X$ and $p_2\stackrel{?}{=} Y$,
which can, and actually will be further corrected by a factor,
which is unity at the pole at $t=0$. A posteriori, we will put this factor equal to $Y^2/\tilde X^3$ (see below).
Then at $\tau=i\infty$, one would get
\be
\frac{c}{1-t}\stackrel{?}{=}x&\hbox{or, after correction,}&x\cdot {y^2\over x^3}={y^2\over x^2} \nn \\
\frac{c^2}{(1-t^2)}\stackrel{?}{=}y&\hbox{or, after correction,}&y\cdot{y^2\over x^3}={y^3\over x^3}
\ee
Looking at (\ref{xyvsz}) we see that the better choice is
\be
\frac{c}{1-t} {=}x+\frac{\pi^2}{3}&\hbox{or, after correction,}&{y^2\over\Big(x+\frac{\pi^2}{3}\Big)^2} \nn \\
\frac{c^2}{(1-t^2)}\stackrel{?}{=}y&\hbox{or, after correction,}&{y^3\over \Big(x+\frac{\pi^2}{3}\Big)^3}
\label{tviaxy}
\ee
This, indeed, corresponds to the choice (\ref{xy}) (up to multiplication of $p_k^\Delta$ with $2\pi$).

We can now express the r.h.s. through $x$ and $y$ --
to obtain rational functions of these variables,--
and then substitute $x\longrightarrow X$, $y\rightarrow Y$.
This will provide elliptic functions with the right asymptotics
in the trigonometric limit at $\tau=i\infty$.

From (\ref{tviaxy}) we get
\be
t = \frac{ y -(x+\frac{\pi^2}{3})}{ y +(x+\frac{\pi^2}{3})}\nn\\
c={(1+t)^2\over 1-t}
\ee
so that the elliptic version is
\be
t = \frac{ Y -\tilde X }{ Y +\tilde X}
\ee
with $\tilde X:=X+\eta/\omega$.

This implies the following expression for elliptic locus $\Delta^{\rm ell}$:
\be
\boxed{
p_k^{\Delta^{\rm ell}} = \frac{\left(\frac{2Y^2}{\tilde X}\right)^k}{(Y+\tilde X)^k-(Y-\tilde X)^k}
}
\ee
in particular
\be
p_1^{\Delta^{\rm ell}} = \frac{Y^2}{\tilde X^2} \ \stackrel{w\rightarrow 0}{\longrightarrow} \
\frac{(1+t)^2}{(1-t)^2} \nn \\
p_2^{\Delta^{\rm ell}} = \frac{Y^3}{\tilde X^3} \ \stackrel{w\rightarrow 0}{\longrightarrow} \
\frac{(1+t)^4}{(1-t^2)(1-t)^2}
\nn \\
\nn\\
\ldots\nn\\
\nn\\
p_k^{\Delta^{\rm ell}} \ \stackrel{w\rightarrow 0}{\longrightarrow} \
\frac{(1+t)^{2k}}{(1-t^k)(1-t)^k}
\ee
%???One can further rescale $c \longrightarrow c \frac{Y^2}{\tilde X^3}$
%to get closer to the original expectation $p_1=\tilde X$,
%still $p_2$ would remain considerably different from $Y$.

Remarkably, Schur functions are spectacularly simple on this locus: they are
fully factorized, and are represented as a product over diagram of the hook length function:
\be\label{20}
\boxed{
\begin{array}{c}
\chi_R\left(p_k^{\Delta^{\rm ell}} \right)=\displaystyle{\left(\frac{2Y^2}{\tilde X}\right)^{|R|}{(Y-\tilde X)^{\nu_R/2}(Y+\tilde X)^{\nu'_R/2}\over\prod_{i,j\in R}\left[
(Y+\tilde X)^{h_{i,j}}-(Y-\tilde X)^{h_{i,j}}\right]}}\cr
\cr
\chi_R\left(\Big[1+(-1)^k\Big]p_k^{\Delta^{\rm ell}} \right)=\displaystyle{\left(\frac{2Y^2}{\tilde X}\right)^{|R|}{{\cal C}_R(Y-\tilde X)^{\mu_R/2}(Y+\tilde X)^{\mu'_R/2}\over\prod_{i,j\in R:\ h_{i,j}\in
\mathbb{Z}_{2n}}\left[
(Y+\tilde X)^{h_{i,j}}-(Y-\tilde X)^{h_{i,j}}\right]}}
\end{array}
}
\ee
Here $\nu'_R:=\nu_{R^\vee}$, $\mu'_R:=\mu_{R^\vee}$ and $^\vee$ means conjugation of the Young diagram.

\section{On the structure of $\mu_R$}

For curiosity, we add a few words about an amusing structure of the function $\mu_R$.

First of all, sometime $\chi_R\left(\Big[1+(-1)^k\Big]p_k^{\Delta^{\rm ell}} \right) \sim {\cal C}_R = 0$,
and this set of diagrams $R$ does not depend on the $t$-deformation,
i.e. is inherited from the vanishing condition of correlators in the original
Gaussian Hermitian model.
For Young diagrams $R$ with up to four lines, it is controlled by the parities of the
sums $R_2+R_3$ and $R_3+R_4$: whenever both (or, equivalently, $R_1+R_2$ and $R_1+R_4$) are odd,
${\cal C}_R$ and thus the entire average of the Schur/Macdonald character vanishes.
For four and six lines, one needs the parity of the sum
$${\cal C}_R = 0 \ \ \Longleftrightarrow \ \
\sum_{i<i'}^3 (R_{2i}-R_{2i'-1})(R_{2i-1}+R_{2i})(R_{2i'-1}+R_{2i'}) \ = \ {\rm odd}$$
for eight and ten lines
{\footnotesize
$${\cal C}_R = 0 \ \ \Longleftrightarrow \ \
\sum_{i<i'}^5 (R_{2i}-R_{2i'-1})(R_{2i-1}+R_{2i})(R_{2i'-1}+R_{2i'}) +
$$ $$
+ \sum_{i<i'<i''<i'''}^4 (R_{2i}-R_{2i'-1})(R_{2i'}-R_{2i''-1})(R_{2i''}-R_{2i'''-1})
(R_{2i-1}+R_{2i})(R_{2i'-1}+R_{2i'})(R_{2i''-1}+R_{2i''})(R_{2i'''-1}+R_{2i'''})\ = \ {\rm odd}$$
}

\noindent
Starting from twelve for the even and already from seven for the odd number of lines,
one needs further corrections, which are not just a naive continuation of
the formulas above.
Note that the vanishing criterium for ${\cal C}_R$ has nothing to do with the
elliptic deformation, it is actually the same already for the ordinary
Gaussian Hermitian model
$$
{\cal C}_R = 0 \ \Longleftrightarrow\  \chi_R\{ p_k=\delta_{k_,2}\} = 0
$$
%for the Schur functions $\chi_R\{p_k\}$.
When ${\cal C}_R$ vanishes, $\mu_R$ is not well-defined.

Next, when $\mu_R$ is defined, it is very similar to $\nu_R=\sum_i (i-1)R_i$,
roughly speaking, $\mu_R\approx \sum_i  (i-1)(R_{2i-1}+R_{2i})$.
A little more accurately, for odd $i-1$ the item $(i-1)(R_{2i-1}+R_{2i})$, when odd,
is substituted by the nearest even integer from above.
But this is not the end of the story, there is still an additional correction:
say, in the case of four lines, the full answer is
$$\mu_R = \sum_{i=1}^2 (i-1)(R_{2i-1}+R_{2i}) + (R_2-R_3+\boxed{1})\cdot\delta_{R_3+R_4,{\rm odd}}$$
where unity in the box is responsible for the above-mentioned substitution
of $R_{3}-R_{4}$ by the nearest even integer.
An exact correction for arbitrary number of lines remains to be found.

Note that expression for $\mu_R$ is also unrelated to ellipticity:
it arises at the level of $t$ deformation, eq.(\ref{4}):
if (\ref{20}) is already known, then the indices $\nu_R$ and $\mu_R$ remain intact
under the further elliptic deformation to (\ref{20}).

\section{Conclusion}

In this paper, we make a somewhat risky suggestion that the peculiar
$p_k^{\Delta_n}$ loci allow an additional elliptic deformation.
If true, this can serve as a first evidence that elliptic matrix models
can be defined with a triple of deformation parameters $q$, $t$, $w:=e^{2\pi i\tau}$,
as  implied  by the elliptic DIM theory in \cite{AKMMgns2}.
We make an explicit suggestion for the  deformation of $p_k^{\Delta_n}$
and consider as a positive evidence of its success
the drastic simplification of
Schur functions on this $p_k^{\Delta^{\rm ell}}$.
The argument is, however, not fully consistent and continues to
rely upon certain guesses.
In particular, there is no clear way to forbid triangular transformations
in the set of elliptic functions with poles of degree $k+1$
(one could add functions with lower poles to our preferred basis).
However, such transformations away from our suggested choice
seem to break the complete factorization of the Schur functions.
We hope that further work will make our guesses more reliable,
what will open a way to consideration
of the full-fledged elliptic GNS functions
from \cite{AKMMgns1,AKMMgns2} on the $p_k^{\Delta^{\rm ell}}$ loci,
which should lead
to an explicit construction of superintegrable elliptic matrix model.

\section*{Acknowledgements}

This work was supported by the Russian Science Foundation (Grant No.20-12-00195).

\end{document}